# Transition between Thin Film Boiling and Evaporation on Nanoporous Membranes Near the Kinetic Limit

Qingyang Wang [a], Yang Shi [a,b], Renkun Chen [a,*]

[a] Department of Mechanical and Aerospace Engineering, University of California, San Diego, La Jolla, California 92093-0411, United States

[b] School of Mechanical Engineering and Automation, Harbin Institute of Technology (Shenzhen), Shenzhen 518000, China

* Email: rkchen@ucsd.edu

**Abstract**

Nanoporous structures including single nanopores and nanoporous membranes have been utilized as a platform to study fundamental liquid-vapor phase change heat transfer (PCHT) processes as well as a promising candidate for high flux heat dissipation. Previously, we implemented nanoporous membranes to support a thin liquid film for boiling, which was termed "thin film boiling", and realized high heat transfer performance. Besides thin film boiling, thin film evaporation through nanoporous structures have also been demonstrated to achieve high heat flux, but these two mechanisms are usually considered two mutually exclusive regimes operated under vastly different conditions, and the factors dictating how close the PCHT process is to the kinetic limit are elusive. In this work, we utilized a unique transition between thin film boiling and evaporation through nanoporous membranes to clarify the factors determining the heat flux and heat transfer coefficient (HTC) with respect to the kinetic limit conditions. We unambiguously showed the controllable transition from boiling to evaporation, when the liquid receded into the nanopores and provided additional driving force from capillary pumping sustained in the nanoscale pores. We showed that this transition is universal and can be understood from a simple fluid transport model for all the four types of fluids we studied, which cover a wide span of surface tension (water, ethanol, IPA, FC-72). More importantly, PCHT conditions at the transition points between boiling and evaporation were close to those of the kinetic limit of all these fluids. However, further increase of the heat flux beyond the transition points led to decreasing HTC and deviation from the kinetic limit, which can be attributed to the increasing vapor resistance in the vapor space and inside the nanopores. This increasing vapor resistance was also confirmed by experiments on IPA with different vapor pressures. Our work could shed light on PCHT on nanoporous structures with respect to the kinetic limit, and could advance the development of high heat-flux heat dissipation devices, especially using dielectric fluids.

**Keywords:** phase change heat transfer, thin film boiling, thin film evaporation, nanoporous membrane, kinetic limit

**Nomenclature**

| | |
|---|---|
| $c_p$ | liquid specific heat capacity, [J kg$^{-1}$ K$^{-1}$] |
| $D_h$ | average hydraulic diameter of the pores, [m] |
| $h_{fg}$ | latent heat of vaporization, [J kg$^{-1}$] |
| $k_l$ | liquid thermal conductivity, [W m$^{-1}$ K$^{-1}$] |
| $L$ | length of the pore, equals to the membrane thickness, [m] |
| $P_c$ | capillary pressure, [Pa] |
| $P_L$ | liquid pressure, [Pa] |
| $P_s$ | saturated vapor pressure at temperature $T_s$, [Pa] |
| $P_V$ | vapor pressure, [Pa] |
| $\Delta P$ | the driving pressure difference for evaporation, equals to $(P_s - P_V)$, [Pa] |
| $q''$ | heat flux, [W m$^{-2}$] |
| $q''_t$ | theoretical maximum heat flux, [W m$^{-2}$] |
| $\overline{q''}$ | normalized heat flux, [-] |
| $R$ | gas constant, [J kg$^{-1}$ K$^{-1}$] |
| $r_p$ | pore radius, [m] |
| $T_L$ | liquid reservoir temperature, [°C] |
| $T_s$ | membrane temperature, [°C] |
| $T_V$ | saturated vapor temperature, [°C] |

*Greek symbols*

| | |
|---|---|
| $\eta$ | membrane porosity, [-] |
| $\mu$ | liquid viscosity, [Pa s] |
| $\Pi$ | liquid property factor normalized by water, defined in Eq. (2), [-] |
| $\rho_l$ | liquid density, [kg m$^{-3}$] |
| $\rho_{v,s}$ | saturated vapor density at $T_s$, [kg m$^{-3}$] |
| $\sigma$ | accommodation coefficient, [-] |

*Abbreviation*

CHF    critical heat flux

HK    Hertz-Knudsen

HTC    heat transfer coefficient

IPA    isopropyl alcohol

PCHT    phase change heat transfer

# 1. Introduction

The utilization of the latent heat of vaporization in liquid-vapor phase change heat transfer (PCHT) allows the transport of a large amount of thermal energy efficiently. As such, liquid-vapor phase change heat transfer has been used extensively for a variety of applications, including power generation, water desalination, and thermal management [1-4]. Many of these applications require high heat flux dissipation (>100 W cm$^{-2}$) with low superheat, which becomes even more demanding for devices with high power density, such as light emitting diode, high performance CPUs and GPUs, laser devices, and electric vehicles, where the efficient removal of heat is of critical importance for the device performance and durability. To address this challenge, numerous studies have been carried out in various types of PCHT schemes, including two distinct modes, boiling and evaporation, to achieve high heat flux and low superheat for high heat transfer capability and efficiency.

There are mainly two types of boiling studied in the literature: pool boiling and flow boiling. In pool boiling, a stagnant bulk pool of liquid is heated up and bubbles are generated on the heater surfaces. The bubble behaviors dictate the heat transfer performance. The critical heat flux (CHF) in pool boiling represents the maximum heat flux that can be dissipated under safe operation conditions, which is usually on the order of 100 W cm$^{-2}$ for a flat surface for water at the atmospheric pressure. Extensive efforts have been made to enhance the CHF and decrease the wall superheat using various strategies, including managing the bubble nucleation sites [5-7], increasing the surface wettability/capillarity using micro/nanostructures (which usually also increases the nucleation site density and provides a fin effect with an enhanced heat transfer area) [8, 9], increasing the contact line length [10, 11], providing separate liquid-vapor pathways for enhanced macroconvection [12, 13], preventing bubble coalescence by pinning the contact line [14], and combination of multiple mechanisms stated above. Significant CHF enhancement has been demonstrated up to ~400 W cm$^{-2}$ [15], but the CHF is still relatively low compared with flow boiling and other new configurations using water [16, 17]. Moreover, the CHF of pool boiling for nonaqueous fluids is usually significantly lower than that of water due to the difference in the thermophysical properties, such as the latent heat of vaporization. For example, CHF values were less than 30 W cm$^{-2}$ for FC-72 [18] or HFE-7100 [19], as the latent heat of these fluids is more than one order of magnitude lower than that of water. In addition, for organic solvents and dielectric fluids that are desirable for cooling of electronic devices, common approaches to manipulate the contact angles for water would not work.

Flow boiling can significant increase the CHF and heat transfer coefficient (HTC) by forcing the liquid to flow along the heated surfaces, and have thus been intensively studied in the past and widely applied for cooling in various systems such as cooling of nuclear reactors, which demands superior stability and reliability for safety. The use of microchannels with various micro- and/or nanostructures [20-22] can significantly reduce the conductive resistance for heat transfer in liquid, promote bubble nucleation, and delay the occurrence of film boiling by enhancing liquid rewetting. However, the large volume expansion of phase change could lead to flow instability inside the microchannels and a pre-mature CHF if not properly designed [23]. For dielectric fluids, it is even more challenging to enhance the CHF due to their poor thermophysical properties [24]. High mass flow rate, subcooled liquid, and delicate structural design have been investigated to achieve higher CHF [25, 26].

As another main mode of phase change heat transfer, evaporation from a thin liquid film has been used in heat pipes and vapor chambers. However, due to the intrinsic low thermal conductivity of the liquid, the majority of heat transfer takes place only at the so-called "thin film evaporating region" of the liquid-vapor interface near the triple-phase contact line [27]. This thin film area only occupies a small portion of the entire interface, and therefore the heat flux of evaporation is low, which limits its application in high heat flux cooling. Micro/nanostructures can provide efficient heat conduction pathways, thereby increase the effective thin film area and decrease the characteristic length for heat transfer in liquid. Besides, they can also provide large capillary force to continuously pump liquid to the interface for vaporization. Thus, they are extensively applied in evaporation studies to minimize the thermal resistance of the system and enhance liquid supply [28-31]. However, these micro/nanostructures, usually with a porous geometry, induce large viscous resistance for fluid transport to the interface, which is dictated by the pore size and coupled with the pumping capillary force. Recently, nanoporous membrane configuration is proposed for achieving high heat flux evaporation by partially decoupling the viscous drag and the capillary driving force [32]. By supplying liquid from the cross-plane direction of the membrane and reducing the membrane thickness, the liquid transport flow length is decreased to lower the viscous drag without significantly affecting the heat transfer and capillary force. Enhanced heat transfer performances have been predicted [33, 34] and demonstrated [35-37].

To achieve high heat flux PCHT, the theoretical limit of the phase change process is considered to be the kinetic limit, which, according to the Hertz-Knudsen (HK) or Schrage formula, essentially relates the maximum vapor mass flux leaving the liquid-vapor interface to the speed of sound. Recent work by Lu et al. [38] and Li et al. [39] showed evaporation near the kinetic limit in nanoporous membranes and single pores, respectively, thus suggesting the prospect of achieving high CHF and high HTC. Nevertheless, the near kinetic limit processes were only observed with low vapor pressure (or low Mach number), thus the attained vapor velocity and eventually the heat flux were not very high, about 340 W cm$^{-2}$ in Lu et al. [38] and 294 W cm$^{-2}$ in Li e al. [39].

This leads us to ask whether or not the near kinetic limit behavior can be universally observed, for example, across different fluids and more importantly, feasible under conditions with larger vapor pressure or Mach number that would result in higher absolute heat flux. Answering these questions would help us better understand and eventually mitigate the factors preventing the PCHT processes from reaching the theoretical limit. In this study, we systematically controlled the experimental conditions such that the PCHT can be tuned to be close to or far away from the kinetic limit. This level of control was achieved by leveraging the concept of "thin film boiling" on nanoporous membranes recently demonstrated by us, which showed high CHF and HTC for boiling due to the small liquid thickness (e.g., 1.8 kW cm$^{-2}$ on 8.4 mm$^2$ area was demonstrated using water as the working fluid [40], and 20 W cm$^{-2}$ K$^{-1}$ on 0.5 cm$^2$ area was demonstrated using ethanol [41]). In this thin film boiling regime, liquid flows through the membrane and forms a thin liquid film for boiling, as shown in Figure 1a. Both the bubble growth and departure are more efficient compared with boiling from a thick liquid pool due to the small thickness of the liquid film. Meanwhile, liquid supply through the porous membrane provides separated liquid-vapor pathways for enhanced liquid rewetting. Therefore, thin film boiling resulted in a significant heat transfer enhancement compared with pool boiling. In this work, we found that the thin film boiling phenomena are universally observed on multiple fluids with vastly different surface tension values, including water, ethanol, isopropyl alcohol (IPA), and FC-72. More importantly, for these fluids, upon further increase of the heat flux, the thin film boiling regime was transitioned to the capillary-driven pore-level evaporation regime when the liquid layer on top of the membrane was depleted by the high heat flux. The heat flux and HTC at the transition points were close to those of the kinetic limit owing to the unique configuration of our experiment which renders the interfacial resistance being the dominant resistance. However, further increase of the heat flux beyond the transition points led to decreasing HTC and deviation from the kinetic limit, which can be attributed to the increasing vapor resistance in the vapor space and inside the nanopores. This increasing vapor resistance was also confirmed by experiments on IPA with different vapor pressures. Our work could lead to a better understanding of the underlying mechanisms of PCHT on nanoporous structures and will benefit the applications such as the design of high heat flux heat sink and thermal devices for thermal management and thermal regulation.

## 2. Experimental

The experimental system, as shown Supplementary Information Figure S2, and the data reduction method used in this work have been employed in our previous studies [40, 41]. Here we summarize the main features of the system for completeness and readability.

There were two chambers in the system, each with a controlled pressure in a specific experiment. The pressure in the liquid chamber, $P_L$, was fixed at a preset value before and during each experiment by adjusting the air pressure inside the chamber with either an air compressor (for above-atmospheric pressure) or a vacuum pump (for sub-atmospheric pressure). The pressure in the vapor chamber ($P_V$) was constantly monitored using a digital pressure gauge and maintained at the preset values using an adjustable valve and a vacuum pump with a PID control loop. The pressure difference between the liquid chamber and the vapor chamber pushed the liquid through the nanoporous membrane, with the excess amount of liquid overflowed to the side of the membrane and eventually reached the bottom of the vapor chamber.

We used nanoporous anodized aluminum oxide (AAO) (Whatman 6809-6022) as the nanoporous membrane. The membrane was epoxy-bonded onto an acrylic sample holder with a 0.71×0.71 cm$^2$ (0.5 cm$^2$ area) hole for liquid flow. A thin Pt film (~70 nm) was deposited on the membrane using magnetron sputtering to serve as the heater and resistance temperature detector simultaneously. In the current work, we implemented a much thinner shadow mask compared to our previous work [40] for Pt deposition, which resulted in a more uniform thickness of the Pt layer to reduce the non-uniformity of heating. Two thick Cu contact pads (~2 μm) were also deposited to allow electrical connection. A 10 nm thick Cr layer was deposited and served as adhesion layer before both the Pt and Cu deposition processes. The size and location of the Pt heater matches the square opening on the sample holder, such that the area with liquid flow matches the area for the heat flux supply. After the deposition of the Pt heater and the Cu contact pads, the sample holder attached with the AAO membrane was assembled onto a custom-made liquid supply channel using screws and O-ring sealing (see Supplementary Information Figure S2d). Four different fluids were used in the experiments: water, ethanol, IPA, and FC-72. De-ionized (DI) water was obtained from a laboratory Milli-Q system. Ethanol (Koptec Pure Ethanol 200 Proof), IPA (ThermoFisher Scientific, HPLC Grade, purity ≥99.9%), and FC-72 (3M Fluorinert) were procured from various vendors. Heat flux was supplied by a direct current (DC) power source to the Pt thin film heater. The voltage drop and the current applied on the Pt heater were measured using two separate digital multimeters. The product of the measured voltage and current was divided by the sample heated area to calculate the heat flux, while the ratio between the measured voltage and current represented the resistance of the heater and was used to obtain the sample temperature using the pre-calibrated temperature coefficient of resistivity (TCR) of the Pt film. Detailed sample preparation procedure, experimental procedure, data reduction method, and the uncertainty analysis can be found in our previous work [40].

## 3. Results and Discussion

### 3.1. PCHT experiments and heat transfer curves

Figure 1a shows the schematic for thin film boiling as we recently reported [40, 41]. The concept is to supply the liquid through a porous membrane and form a thin liquid film on top of the membrane, and the thermal resistance of conduction inside liquid phase is effectively reduced due to the small thickness of the liquid film. In the thin film boiling configuration, the liquid floods atop the membrane so there is no capillary force from the nanopores in the membrane. As such, the liquid supply through the membrane is driven by the pressure difference across the membrane, and the flow rate associated with this liquid supply governs the maximum heat flux in this regime, as we observed previously in water [40]. By comparing the work between evaporation studied by Wang and coworkers [32, 36] and thin film boiling by us [40, 41] on nanoporous membranes, as well as recent work on evaporation from nanochannels and nanopores by Duan and coworkers [39, 42], it is evident that both heat transfer mechanisms could take place on nanoporous structures. However, it remains unclear what are the conditions leading to boiling or evaporation and if a transition between these two regimes is possible. We note that as the liquid on top of the membrane is dried out, it could recede into the nanopores and form liquid-vapor menisci, as shown in Figure 1b-c. If this occurs, the capillary pressure inside the pores can provide extra liquid flow rate, exceeding the amount driven by the external liquid pumping pressure. This transition from thin film boiling to pore-level thin film evaporation was observed experimentally by us on an organic solvent, ethanol [41], which led to a heat flux considerably higher than what was predicted based on the liquid pressure driven flow rate in the thin film boiling regime. However, the same phenomenon was not observed in water [40]. Therefore, we first set out to test this transition on a variety of fluids, including water, ethanol, IPA, and FC-72.

We tested the heat transfer using four fluids at various liquid pressures and a fixed vapor pressure for each fluid. Figure 2 shows the heat transfer curves of four different working fluids: water, ethanol, IPA, and FC-72. For the curves shown in Figure 2, the vapor pressure $P_V$ for water, ethanol, IPA, and FC-72 were fixed at 2.3, 5.9, 4.4, and 25.2 kPa, respectively. These pressures are the saturation pressures of 20 °C vapor for the respective fluid types. The liquid pressure $P_L$ was varied for each fluid.

It can be seen from Figure 2 that for all of the fluids tested in this work, the curves display a small-slope regime at low heat fluxes, which represents the well-known pool boiling regime, as shown previously [40, 41]. This is because at low heat fluxes, the liquid flow across the membrane driven by the pressure difference ($P_L - P_V$) was much larger than the vaporization flux, which resulted in a thick liquid puddle on top of the membrane and led to heat transfer behavior similar to pool boiling. When the liquid film is much thicker than the thermal boundary layer thickness, which is tens to a few hundred μm for these fluids using the scaling analysis [40], the boiling lies in the pool boiling regime, where the bubble departure diameter is ~mm and the bubble behavior is similar to pool boiling, as shown in Supplementary Information Section S5. The bubble formation also indicates that the beginning of the curves with small (positive) slopes are already in the nucleate boiling regime instead of natural convection.

It is worth mentioning that due to the more uniform thickness of the deposited Pt layer (as described in Section 2), we obtained higher heat flux and HTC in the pool boiling regime than those obtained in our previous work [40] for the same fluid (i.e., water). The heat transfer curves are still largely following the trend predicted by the Rohenow correlation as shown in Figure 2d, but with higher HTC, which can be attributed to the finite thickness of the liquid film as well as the nanoporous morphology of the membrane that could possibly offer more nucleate sites [8, 13].

As the heat flux increases, the slope of the curves becomes negative, which is what we referred to as the "thin film boiling" regime. In this regime, the increase of heat flux caused the reduction of the liquid layer thickness on top of the membrane, which led to a smaller thermal resistance for bubble growth and smaller bubble departure diameter, and consequently the decrease of superheat. This regime gave rise to the interesting negative slopes in the boiling curves, as shown in all the four fluids in Figure 2. It is worth mentioning that negative slopes in boiling curves have been reported before for pool boiling [8, 43], which were attributed to the activation of small nucleation sites and thus exhibited hysteresis when increasing and decreasing the heat flux. On the contrary, in this thin film boiling regime, the negative slope was fully reversible and showed no hysteresis [41] since it was caused by the liquid layer thickness change. In this regime, as the liquid layer thickness decreased, the bubble size also reduced to tens to a few hundred μm.

Different trends can be observed between the curves for water and the curves for other fluids at the end of the thin film boiling regime. In the case for water, after the termination of the negative slope portion, the curves ended, and no higher heat flux was achieved in the experiments. On the other hand, in the curves for low surface tension fluids, another portion with a positive slope emerged. The extra positive slope in the curves for low surface tension fluids represented the pore-level thin film evaporation regime, and the point where the slope of the curve turned from negative to positive was considered the "transition point", namely, indicating the transition from thin film boiling to pore-level evaporation. When the vaporization flux is balanced by the maximum liquid flux that can be provided by the pressure difference ($P_L - P_V$) across the membrane, liquid layer on top of the membrane would dry out, which occurred at the transition points. Any further increase of the heat flux would cause the liquid-vapor interface to recede inside the pores and generate capillary pressure for sustaining the higher liquid flux until the eventual CHF was reached. In this capillary-aided evaporation regime, as more vapor needs to be removed from the interface with increasing heat flux, the vapor removal resistance becomes increasingly dominant, resulting in the positive slope, as will be discussed in Sections 3.3 and 3.4.

The absence of this pore-level evaporation regime in the water experiments was probably caused by the poor wettability of water on the heater surface, such that any local dry-out spots cannot be quickly replenished by lateral spreading of water on the heater surface. When the thin film boiling regime ended (marked by the stars in Figure 2, including the transition points in Figure 2a-c and CHFs in Figure 2d), the non-uniform heating from the Pt film would inevitably cause some local spots having larger heat fluxes relative to the rest area of the heater, for example, due to variations in the pore size and the Pt film thickness. These hot spots could cause pre-mature CHF before the evaporation regime was established in water. On the contrary, this pre-mature failure caused by local hot spots can be prevented for low surface tension liquids due to their better lateral liquid spreading capability (with contact angle close to zero, compared with water whose contact angle is ~60° (Ref. [41])). The liquid spreading due to the wetting of liquid on the membranes can help dissipate higher heat flux [39] and mitigate the hotspots, thus enabling the formation of stable pore-level evaporation regime.

The maximum heat flux dissipated in the thin film boiling regime, which corresponds to the CHF in the experiments with water and the heat fluxes at the transition points in experiments with other fluids, highly depends on both the experimental condition (liquid pressure and vapor pressure) and the working fluid used, which will be discussed in the following sections.

### 3.2. Universal thin film boiling behaviors on multiple fluids

In order to predict the maximum heat flux that can be sustained by thin film boiling, a model based on fluid transport was developed. As shown in Figure 2, the maximum heat flux within the *thin film boiling* regime corresponds to the CHFs in the water experiments and the transition heat flux in the experiments with other fluids. This maximum heat flux in thin film boiling was achieved when the liquid flow rate driven by the pressure difference between the liquid and vapor chambers was balanced with the vapor flux from the boiling process. Therefore, the maximum heat flux can be estimated as [40]

$$q_t'' = \frac{D_h^2 (P_L - P_V)}{32 \mu L} \rho_l [h_{fg} + c_p (T_s - T_L)] \eta \qquad (1)$$

if assuming circular pore geometry in the AAO. Here, $D_h$ is the average hydraulic diameter of the pores, $L$ is the membrane thickness, $\eta$ is the membrane porosity, $T_s$ and $T_L$ are the wall temperature and liquid reservoir temperature, respectively, $\mu$, $\rho_l$, $h_{fg}$, and $c_p$ are the viscosity, density, latent heat, and specific heat capacity of the working fluid, respectively. In this equation, the viscosity of the liquid $\mu$ is highly temperature dependent. By using viscosity values at different temperatures within the experimental temperature range, the calculated heat flux can vary by several times. We choose the wall temperature $T_s$, which was also the sample surface temperature measured from the Pt thin film heater, to extract the viscosity, as the fluid inside the AAO was estimated to be at the same temperature as $T_s$ based on our heat transfer analysis (see Supplementary Information Section S1).

In order to better reflect the effect of the thermophysical properties of the liquid on thin film boiling performance, we defined a normalized liquid property factor $\Pi$ as [44]

$$\Pi = \frac{\rho_l h_{fg}}{\mu} \bigg/ \frac{\rho_{water} h_{fg,water}}{\mu_{water}} \qquad (2)$$

$\Pi$ is normalized using water properties and represents the ability to achieve high heat flux in thin film boiling according to Eq. (1). The experimental maximum heat fluxes achieved by thin film boiling for different liquids (the CHF for water, including the data reported in Ref. [40], and the transition heat flux for other fluids in Figure 2) are divided by the respective liquid property factor, and are plotted against the pressure difference as shown in Figure 3. The properties used for calculation of the $\Pi$ factor are all chosen at 20 °C for simplicity, since the wall temperature at the maximum heat flux points of thin film boiling varies for different liquids.

It can be seen from Figure 3 that the $\Pi$-normalized maximum thin film boiling heat flux as a function of pressure difference for various fluids are converged into one straight line from the origin. In another word, the normalized maximum heat flux has a linear relationship with the pressure difference, and the slopes of this linear relationship for different fluids are almost the same. The calculated results from Eq. (1) for water with different superheat are also shown in Figure 3. The agreement between the normalized heat flux and the calculated transition further confirmed the validity of the model. We notice that there is still certain mismatch in the results, and the convergence of the data points are not perfect. For example, the data points for ethanol are slightly below the points for other fluids. This can be attributed to the fact that we used 20 °C to evaluate all the liquid properties. In the real experiments, the liquids were not at exact 20 °C, and the thermophysical properties of different fluid could have different extent of temperature dependence, which may have resulted in errors in using the liquid property factor $\Pi$. Nevertheless, we can still use the liquid property factor for the purpose of fluid selection and heat flux prediction in the thin film boiling configuration. And the fact that these four fluids with vastly different surface tension values all show the same behaviors strongly demonstrate the universality for the thin film boiling scheme. This is especially important for organic solvents and dielectric fluids, as they tend to have much lower boiling CHFs due to their low latent heat and are difficult to engineer due to their low surface tension (virtually all the surfaces are highly wettable for these fluids).

### 3.3. Near kinetic limit at the onset of pore-level thin film evaporation

The evaporative heat flux $q''$ can be normalized using the maximum heat flux equation [45] as $\overline{q''} = q''/(\rho_{v,s} h_{fg} \sqrt{\frac{RT_s}{2\pi}})$ where $R$ is the gas constant, $T_s$ is the membrane surface temperature which also represents the liquid temperature, and $\rho_{v,s}$ is the density of saturated vapor at $T_s$. The experimental results at the transition points between the boiling and evaporation regimes (i.e., onset of pore-level evaporation or the termination point of thin film boiling regime) shown by stars in Figure 2 for different fluids are normalized and shown in Figure 4, where the horizontal axis is the dimensionless driving potential for evaporation [38] $\Delta P/P_s$ where $\Delta P = P_s - P_V$ and $P_s$ is the saturated pressure at $T_s$. Although this dimensionless driving potential is not directly calculated from superheat, it is closely related to superheat which is calculated as $T_s - T_V$ where $T_V$ is the saturation temperature at $P_V$.

The kinetic limited evaporation heat flux is also calculated using the Hertz-Knudsen equation (HK equation, see Supplementary Information Section S4) and plotted in Figure 4 with different values for the accommodation coefficient ($\sigma$), which represents the ratio of the number of molecules evaporated from the interface to the number of liquid molecules at the interface. Note that this model calculates the heat flux from a flat liquid-vapor interface, corresponding to the case where the liquid is evaporating from the entire membrane surface, which could be true at the transition points since the liquid might spread outside the pores when the interface is pinned at the entrance of the pores [39], while the interface receding inside the pores is more likely to happen with higher heat flux. The experimental results at the onset of the evaporation regime showed near kinetic limit behavior and agree well with the calculated results from the kinetic limit with a reasonable range for the accommodation coefficient range (0.1 to 0.45) [36, 38], which indicate that the heat transfer characteristics for the transition points are close to those of the kinetic limit. This agreement shows that the liquid-vapor interfacial resistance is the dominant one, compared to other possible resistances in the system (such as fluid flow and vapor transport).

In the thin film boiling regime, the thickness of the liquid induces thermal resistance for conduction. Right after the transition happened, the characteristic length scale of the liquid layer became the pore radius $r_p$. That is, the conduction thermal resistance in the liquid phase (scales with $r_p/k_l$, where $k_l$ is the liquid thermal conductivity) became small and resulted in low superheat to sustain the heat flux at the transition point. For example, in the IPA experiment with 101.3 kPa liquid pressure (the purple circles in Figure 2b), the conduction resistance estimated by $r_p/k_l$ is ~7×10$^{-7}$ m$^2$ K W$^{-1}$, which is still orders of magnitude smaller than the overall thermal resistance of the system (~1.6×10$^{-5}$ m$^2$ K W$^{-1}$ achieved at the transition point). Therefore, the liquid thermal conduction resistance is not important in the thin film evaporation regime.

At the pore-level evaporation regime, three possible resistances could limit the highest heat flux that can be dissipated: the kinetic limited interfacial resistance of evaporation, the hydraulic resistance of liquid transport, and the vapor diffusion/advection resistance from interface to the far field in the vapor space. At the onset of evaporation regime, the liquid-vapor interface forms menisci and provides sufficient liquid transport, rendering the liquid transport resistance less important, and the interface is mostly likely either pinned on top of the pores or spread outside the pores, rendering the vapor removal resistance small. Therefore, at these transition conditions, the kinetic limited interfacial resistance becomes the dominant resistance, resulting in the heat transfer characteristics close to those predicted by the kinetic limit.

Due to the near-kinetic-limit condition, PCHT at the transition show high heat flux achieved at small superheat. Figure 5 shows the heat flux vs. HTC plot of the transition points measured in this work for various fluids (stars in Figure 2), along with other experimental results on PCHT of these fluids reported in the literature, showing that the transition points for these fluids possess both high heat flux and high HTC, compared to the same fluids from the previously reported pool boiling and evaporation experiments. In our previous work, a CHF of over 1.8 kW cm$^{-2}$ was achieved for experiments with water, which is close to the previously reported record-high CHF in boiling heat transfer by Moghaddam et al. [17]. In this work, what is perhaps more remarkable is for the other three fluids with lower surface tension, which showed that the heat fluxes and HTC at the transition points are significantly higher than those from the traditional pool boiling or evaporation. For example, CHF of ~60 W cm$^{-2}$ and HTC of 2.7 W cm$^{-2}$ K$^{-1}$ on a plain surface and CHF of ~110 W cm$^{-2}$ and HTC of 6.5 W cm$^{-2}$ K$^{-1}$ on structured surfaces were achieved in pool boiling with ethanol at atmospheric pressure [46], and the CHF decreases with decreasing saturation pressure. In comparison, the transition heat flux of our current work using ethanol is exceeding 200 W cm$^{-2}$ with HTC close to 28 W cm$^{-2}$ K$^{-1}$ with 5.9 kPa saturation pressure. Moreover, the maximum transition heat flux for FC-72 is 75 W cm$^{-2}$ and corresponding HTC is over 9 W cm$^{-2}$ K$^{-1}$, which is significantly higher than the CHF (<30 W cm$^{-2}$) and HTC (<2.3 W cm$^{-2}$ K$^{-1}$) reported in pool boiling literature [18, 47, 48]. More importantly, the significant CHF and HTC enhancement achieved at the transition points, which were universally demonstrated for various fluids, indicates the possibility of using nanoporous membranes for achieving kinetic limited high heat flux PCHT.

### 3.4. Vapor resistances in thin film evaporation regime

For all of the curves for low surface tension liquids, although there were stable thin film evaporation regimes beyond the transition points shown in the previous section, the HTC (or slope of heat flux vs. superheat) decreased and became far lower than that of the transition points as shown in Figure 2. We believe this rapid decrease in HTC is most likely caused by the increasing vapor resistance when the evaporation progressed beyond the transition points. Since the HTC of the evaporation regime is smaller than that of the transition points which is close to the kinetic limit (see Supplementary Information Section S6), there must be another dominant resistance that is much greater than the interfacial resistance in this evaporation regime. While we were unable to directly measure the remaining two resistances (namely, fluid flow and vapor transport) in our experiments, here we carried out additional experiment that suggested the dominant effect of the vapor resistance.

We used IPA as an example, and conducted experiments with various vapor pressures while maintaining the pressure difference across the membrane in a narrow range, as shown in Figure 6a. The vapor pressure was varied from 5 values: 1.1, 2.3, 4.4, 8.2, and 14.3 kPa, corresponding to the saturated IPA vapor temperature of 0, 10, 20, 30, and 40 °C, respectively. Except for the experiments with vapor pressure at 14.3 kPa, the liquid pressure was fixed at atmospheric pressure (101.3 kPa). For the vapor pressure at 14.3 kPa, the liquid pressure was fixed at 111.3 kPa so that the pressure difference across the membrane (97.0 kPa) was close to that of the other experimental conditions (ranging from 93.1 to 100.2 kPa). In Figure 6b, the curves from Figure 6a are truncated at the transition points (between the boiling and evaporation regimes), and are shown as heat flux as a function of superheat.

It can be seen in Figure 6a that all of the curves display the typical profile with three regimes as discussed above, but the experiments under different vapor pressures show different transition heat fluxes and at different wall temperatures. This can be explained again by the fluid transport through the membrane. As discussed before, the transition point shows near kinetic limit behavior with small wall superheat, as seen in Figure 6b. The different saturation pressures with similarly small superheat gave different wall temperatures (which are also fluid temperatures) and thus significantly different fluid viscosities, and with similar driving force for liquid transport ($P_L - P_V$), vastly different transition heat fluxes were obtained. Due to the varying saturation temperature for the curves shown in Figure 6b, the transition points of the curves were not plotted in Figure 3 which is based on the liquid property factor $\Pi$ evaluated at 20 °C. The transition heat flux estimated by Eq. (1) is also shown in Figure 6b as dashed lines with different colors representing corresponding vapor pressure conditions, showing good agreement between the expected transition heat flux and the experimental results.

In Figure 6c, the experimental evaporation regime for the experiments with IPA at various vapor pressures (shown in Figure 6a) are also normalized and plotted along with the HK models.

The experimental heat flux deviates away from the kinetic limit as the driving potential increases (which corresponds to the superheat increase, as shown in Figure S4 of the Supplementary Information), which is likely due to the increasing vapor resistance. When heat flux (and also superheat) is increased during the experiment for the evaporation regime, a higher flux of vapor molecules need to be removed from the interface. Meanwhile, the interface is more likely to recede inside the pores and result in extra resistance for vapor to escape from the pores. Both the increased flux of vapor molecules and the possibility of interface receding increase the vapor diffusion/advection resistance, and thus the experimentally achieved heat flux is significantly smaller than the kinetic limit. The normalized flux is decreasing as heat flux increases, and the CHF achieved is still much smaller than what could be expected (see Supplementary Information Section S2), indicating the increasing dominance of vapor resistance. We also notice that the experimental curves with smaller vapor pressure is closer to the calculated kinetic limit, which can also be attributed to the smaller vapor resistance at the lower pressure (a situation that is similar to the near-kinetic limit behavior at low Mach number, as reported by Lu et al. [38]).

More experiments are carried out following the same conditions shown in Figure 6a, and Figure 6d shows the recorded CHF as a function of the vapor pressure in these experiments with IPA. The pressure difference across the membrane for these experiments are all similar (93.1~100.2 kPa) despite different vapor pressures, but the resulting CHF values show a decreasing trend with increasing vapor pressure, which further proved the dominance of vapor resistance. If a nanoporous membrane with highly uniform pore geometry and heater deposition is used to eliminate the hotspot, and a pure vapor ambient with low pressure is maintained, a much higher heat flux could be possibly realized through thin film evaporation [38].

## 4. Conclusion

In this work, we experimentally studied thin film boiling and evaporation heat transfer of different types of fluids through nanoporous membranes. Transition from thin film boiling to evaporation was observed for the three low surface tension fluids (ethanol, IPA, and FC-72) but not for water, which was attributed to hotspot formation due to membrane nonuniformity. The transition heat flux for these fluids universally agreed with the model prediction. At the transition points, interfacial resistance dominates the heat transfer behavior which resulted in heat flux and HTC close to the kinetic limit. After transition into the pore-level evaporation regime, the vapor resistance became increasingly dominant and eventually limited the achievable CHF. Our work provides a systematic study of PCHT through nanoporous membranes and can benefit the understanding and application of PCHT through nanoporous membranes for thermal management and modulation.

**Acknowledgements**

This work was supported in part by an ACS PRF grant (54109-ND10), UC Solar Institute (UC Multicampus Research Programs and Initiatives grant MR-15-328386), and National Science Foundation (CMMI- 1762560). Y.S. was supported by Chinese Scholarship Council of China.

# References

[1] H.J. Cho, D.J. Preston, Y. Zhu, E.N. Wang, Nanoengineered materials for liquid–vapour phase-change heat transfer, Nature Reviews Materials, 2(2) (2017) 16092.

[2] R. Wen, X. Ma, Y.-C. Lee, R. Yang, Liquid-Vapor Phase-Change Heat Transfer on Functionalized Nanowired Surfaces and Beyond, Joule, 2(11) (2018) 2307-2347.

[3] N.S. Dhillon, J. Buongiorno, K.K. Varanasi, Critical heat flux maxima during boiling crisis on textured surfaces, Nature communications, 6 (2015) 8247.

[4] J. Zeng, Q. Wang, Y. Shi, P. Liu, R. Chen, Osmotic Pumping and Salt Rejection by Polyelectrolyte Hydrogel for Continuous Solar Desalination, Advanced Energy Materials, 9(38) (2019) 1900552.

[5] M.-C. Lu, R. Chen, V. Srinivasan, V.P. Carey, A. Majumdar, Critical heat flux of pool boiling on Si nanowire array-coated surfaces, International Journal of Heat and Mass Transfer, 54(25) (2011) 5359-5367.

[6] Y. Liu, J. Tang, L. Li, Y.N. Shek, D. Xu, Design of Cassie-wetting nucleation sites in pool boiling, International Journal of Heat and Mass Transfer, 132 (2019) 25-33.

[7] Y. Liu, M.-C. Lu, D. Xu, The suppression effect of easy-to-activate nucleation sites on the critical heat flux in pool boiling, International Journal of Thermal Sciences, 129 (2018) 231-237.

[8] R. Chen, M.-C. Lu, V. Srinivasan, Z. Wang, H.H. Cho, A. Majumdar, Nanowires for enhanced boiling heat transfer, Nano letters, 9(2) (2009) 548-553.

[9] S. Shin, G. Choi, B. Rallabandi, D. Lee, D.I. Shim, B.S. Kim, K.M. Kim, H.H. Cho, Enhanced Boiling Heat Transfer using Self-Actuated Nanobimorphs, Nano letters, 18(10) (2018) 6392-6396.

[10] P. Raghupathi, S. Kandlikar, Pool boiling enhancement through contact line augmentation, Applied Physics Letters, 110(20) (2017) 204101.

[11] A. Zou, D.P. Singh, S.C. Maroo, Early evaporation of microlayer for boiling heat transfer enhancement, Langmuir : the ACS journal of surfaces and colloids, 32(42) (2016) 10808-10814.

[12] S.G. Kandlikar, Enhanced Macroconvection Mechanism With Separate Liquid–Vapor Pathways to Improve Pool Boiling Performance, Journal of Heat Transfer, 139(5) (2017) 051501.

[13] R. Wen, Q. Li, W. Wang, B. Latour, C.H. Li, C. Li, Y.-C. Lee, R. Yang, Enhanced bubble nucleation and liquid rewetting for highly efficient boiling heat transfer on two-level hierarchical surfaces with patterned copper nanowire arrays, Nano Energy, 38 (2017) 59-65.

[14] T.P. Allred, J.A. Weibel, S.V. Garimella, Enabling Highly Effective Boiling from Superhydrophobic Surfaces, Physical review letters, 120(17) (2018) 174501.

[15] A. Jaikumar, S.G. Kandlikar, Pool boiling enhancement through bubble induced convective liquid flow in feeder microchannels, Applied Physics Letters, 108(4) (2016) 041604.

[16] J.W. Palko, H. Lee, C. Zhang, T.J. Dusseault, T. Maitra, Y. Won, D.D. Agonafer, J. Moss, F. Houshmand, G. Rong, Extreme Two‐Phase Cooling from Laser‐Etched Diamond and Conformal, Template‐Fabricated Microporous Copper, Advanced Functional Materials, 27(45) (2017).

[17] A. Fazeli, S. Moghaddam, A New Paradigm for Understanding and Enhancing the Critical Heat Flux (CHF) Limit, Scientific reports, 7(1) (2017) 5184.

[18] H. Seo, J.H. Chu, S.-Y. Kwon, I.C. Bang, Pool boiling CHF of reduced graphene oxide, graphene, and SiC-coated surfaces under highly wettable FC-72, International Journal of Heat and Mass Transfer, 82 (2015) 490-502.

[19] S.J. Thiagarajan, R. Yang, C. King, S. Narumanchi, Bubble dynamics and nucleate pool boiling heat transfer on microporous copper surfaces, International Journal of Heat and Mass Transfer, 89 (2015) 1297-1315.

[20] F. Yang, X. Dai, C.-J. Kuo, Y. Peles, J. Khan, C. Li, Enhanced flow boiling in microchannels by self-sustained high frequency two-phase oscillations, International Journal of Heat and Mass Transfer, 58(1-2) (2013) 402-412.

[21] Y. Zhu, D.S. Antao, K.-H. Chu, S. Chen, T.J. Hendricks, T. Zhang, E.N. Wang, Surface Structure Enhanced Microchannel Flow Boiling, Journal of Heat Transfer, 138(9) (2016) 091501.

[22] W. Li, X. Qu, T. Alam, F. Yang, W. Chang, J. Khan, C. Li, Enhanced flow boiling in microchannels through integrating multiple micro-nozzles and reentry microcavities, Applied Physics Letters, 110(1) (2017) 014104.

[23] T. Zhang, Y. Peles, J.T. Wen, T. Tong, J.-Y. Chang, R. Prasher, M.K. Jensen, Analysis and active control of pressure-drop flow instabilities in boiling microchannel systems, International Journal of Heat and Mass Transfer, 53(11-12) (2010) 2347-2360.

[24] J. Ma, W. Li, C. Ren, J.A. Khan, C. Li, Realizing highly coordinated, rapid and sustainable nucleate boiling in microchannels on HFE-7100, International Journal of Heat and Mass Transfer, 133 (2019) 1219-1229.

[25] C. Woodcock, C. Ng'oma, M. Sweet, Y. Wang, Y. Peles, J. Plawsky, Ultra-high heat flux dissipation with Piranha Pin Fins, International Journal of Heat and Mass Transfer, 128 (2019) 504-515.

[26] K.P. Drummond, D. Back, M.D. Sinanis, D.B. Janes, D. Peroulis, J.A. Weibel, S.V. Garimella, Characterization of hierarchical manifold microchannel heat sink arrays under simultaneous background and hotspot heating conditions, International Journal of Heat and Mass Transfer, 126 (2018) 1289-1301.

[27] J. Plawsky, A. Fedorov, S. Garimella, H. Ma, S. Maroo, L. Chen, Y. Nam, Nano-and microstructures for thin-film evaporation—A review, Nanoscale and microscale thermophysical engineering, 18(3) (2014) 251-269.

[28] X. Dai, M. Famouri, A.I. Abdulagatov, R. Yang, Y.-C. Lee, S.M. George, C. Li, Capillary evaporation on micromembrane-enhanced microchannel wicks with atomic layer deposited silica, Applied Physics Letters, 103(15) (2013) 151602.

[29] X. Dai, F. Yang, R. Yang, Y.-C. Lee, C. Li, Micromembrane-enhanced capillary evaporation, International Journal of Heat and Mass Transfer, 64 (2013) 1101-1108.

[30] X. Dai, F. Yang, R. Yang, X. Huang, W.A. Rigdon, X. Li, C. Li, Biphilic nanoporous surfaces enabled exceptional drag reduction and capillary evaporation enhancement, Applied Physics Letters, 105(19) (2014) 191611.

[31] R. Wen, S. Xu, Y.-C. Lee, R. Yang, Capillary-Driven Liquid Film Boiling Heat Transfer on Hybrid Mesh Wicking Structures, Nano Energy,  (2018).

[32] R. Xiao, S.C. Maroo, E.N. Wang, Negative pressures in nanoporous membranes for thin film evaporation, Applied Physics Letters, 102(12) (2013) 123103.

[33] D.F. Hanks, Z. Lu, S. Narayanan, K.R. Bagnall, R. Raj, R. Xiao, R. Enright, E.N. Wang, Nanoporous evaporative device for advanced electronics thermal management, in:  Thermal and Thermomechanical Phenomena in Electronic Systems (ITherm), 2014 IEEE Intersociety Conference on, IEEE, 2014, pp. 290-295.

[34] Z. Lu, S. Narayanan, E.N. Wang, Modeling of Evaporation from Nanopores with Nonequilibrium and Nonlocal Effects, Langmuir : the ACS journal of surfaces and colloids, 31(36) (2015) 9817-9824.

[35] Z. Lu, K.L. Wilke, D.J. Preston, I. Kinefuchi, E. Chang-Davidson, E.N. Wang, An Ultrathin Nanoporous Membrane Evaporator, Nano letters, 17(10) (2017) 6217-6220.

[36] K.L. Wilke, B. Barabadi, Z. Lu, T. Zhang, E.N. Wang, Parametric study of thin film evaporation from nanoporous membranes, Applied Physics Letters, 111(17) (2017) 171603.

[37] D.F. Hanks, Z. Lu, J. Sircar, T.R. Salamon, D.S. Antao, K.R. Bagnall, B. Barabadi, E.N. Wang, Nanoporous membrane device for ultra high heat flux thermal management, Microsystems & Nanoengineering, 4(1) (2018) 1.

[38] Z. Lu, I. Kinefuchi, K.L. Wilke, G. Vaartstra, E.N. Wang, A unified relationship for evaporation kinetics at low Mach

numbers, Nature Communications, 10(1) (2019) 2368.

[39] Y. Li, H. Chen, S. Xiao, M.A. Alibakhshi, C.-W. Lo, M.-C. Lu, C. Duan, Ultrafast Diameter-Dependent Water Evaporation from Nanopores, ACS nano, 13(3) (2019) 3363-3372.

[40] Q. Wang, R. Chen, Ultrahigh Flux Thin Film Boiling Heat Transfer Through Nanoporous Membranes, Nano Letters, 18(5) (2018) 3096-3103.

[41] Q. Wang, R. Chen, Widely tunable thin film boiling heat transfer through nanoporous membranes, Nano Energy, 54 (2018) 297-303.

[42] Y. Li, M.A. Alibakhshi, Y. Zhao, C. Duan, Exploring Ultimate Water Capillary Evaporation in Nanoscale Conduits, Nano Letters,  (2017).

[43] C. Kruse, A. Tsubaki, C. Zuhlke, T. Anderson, D. Alexander, G. Gogos, S. Ndao, Secondary pool boiling effects, Applied Physics Letters, 108(5) (2016) 051602.

[44] Z. Lu, T.R. Salamon, S. Narayanan, K.R. Bagnall, D.F. Hanks, D.S. Antao, B. Barabadi, J. Sircar, M.E. Simon, E.N. Wang, Design and modeling of membrane-based evaporative cooling devices for thermal management of high heat fluxes, IEEE Transactions on Components, Packaging and Manufacturing Technology, 6(7) (2016) 1056-1065.

[45] V.P. Carey, Liquid-vapor Phase-change Phenomena: An Introduction to the Thermophysics of Vaporization and Condensation Processes in Heat Transfert Equipment, Hemisphere publishing corporation, 1992.

[46] A. Kalani, S.G. Kandlikar, Enhanced pool boiling with ethanol at subatmospheric pressures for electronics cooling, Journal of Heat Transfer, 135(11) (2013) 111002.

[47] Z.W. Liu , W.W. Lin , D.J. Lee, X.F. Peng Pool Boiling of FC-72 and HFE-7100, Journal of Heat Transfer, 123(2) (1999) 399-400.

[48] M.S. El-Genk, J.L. Parker, Nucleate boiling of FC-72 and HFE-7100 on porous graphite at different orientations and liquid subcooling, Energy Conversion and Management, 49(4) (2008) 733-750.

[49] K. Wilke, Evaporation from Nanoporous Membranes, Master's thesis, Massachusetts Institute of Technology, Cambridge, 2016.

[50] A. Jaikumar, T. Emery, S. Kandlikar, Interplay between developing flow length and bubble departure diameter during macroconvection enhanced pool boiling, Applied Physics Letters, 112(7) (2018) 071603.

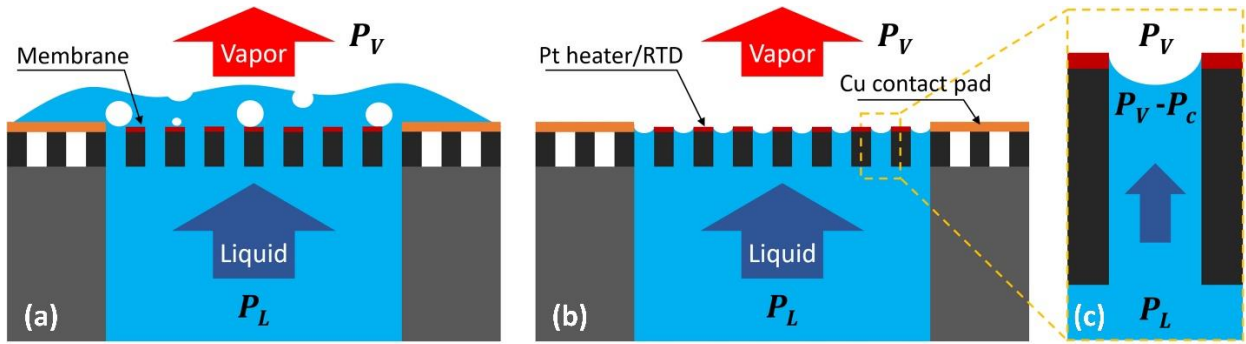

*Figure 1: Schematic illustrations showing (a) thin film boiling, where bubbles are generated on top of the heated nanoporous membranes, (b) pore-level thin film evaporation when the liquid is receded into the pores, and (c) zoom-in schematic of a single pore from (b) showing the capillary pressure aided liquid supply.*

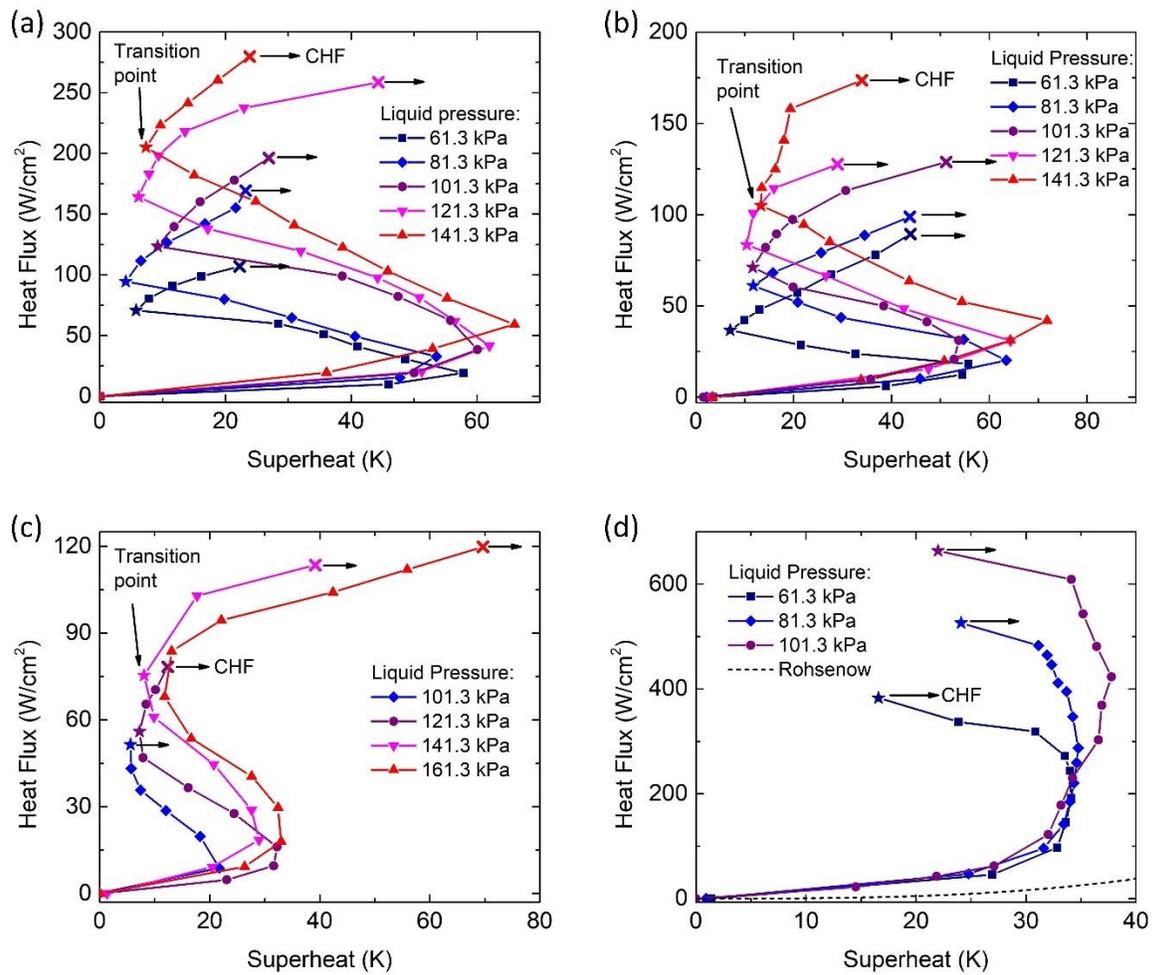

*Figure 2: Heat transfer curves for (a) ethanol, (b) IPA, (c) FC-72, and (d) water at 20 °C saturated vapor with varying liquid pressures. Figure 2a is adapted from the ethanol experiment reported in Ref. [41]. The "transition points" indicate the transition from thin film boiling to thin film evaporation, represented by the stars. Beyond these transition points, there are thin film evaporation regimes in ethanol, IPA, and FC-72, but not in water. The Rohsenow correlation for nucleate boiling under the experimental condition is also plotted in (d), showing that the small slope portion of the curves are close to the Rohsenow model prediction.*

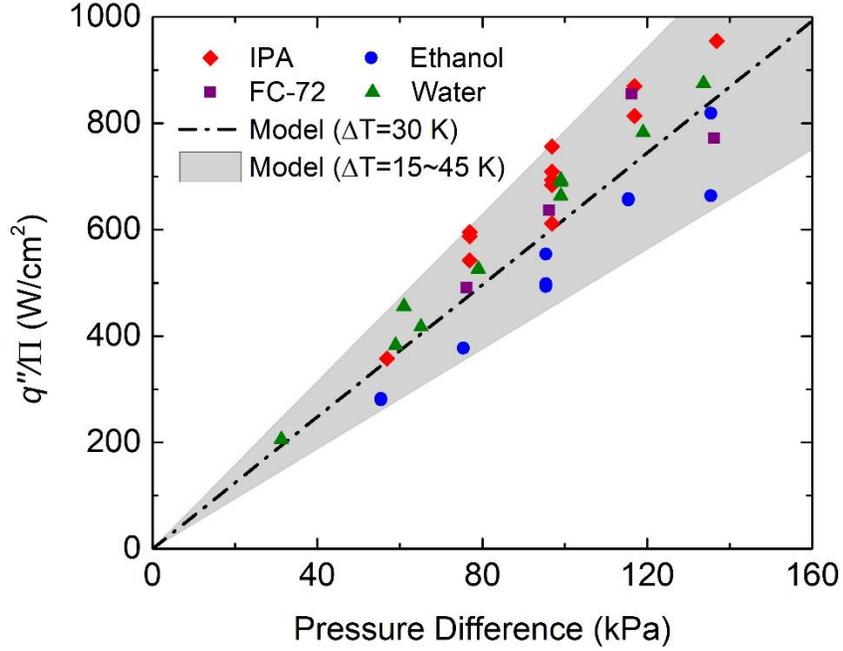

*Figure 3: Maximum heat flux for thin film boiling (shown by the stars in Figure 2, or heat flux at transition points for ethanol, IPA, FC-72, and the CHF for water) divided by the liquid property factor Π (defined in Eq. (2)) as a function of pressure difference for various fluids. Experimental data for water from Ref. [40] are also included. The model of transition for water calculated using Eq. (1) is also shown, where the dash dot line corresponds to a superheat of 30 K and the grey band corresponds to the superheat range of 15~45 K.*

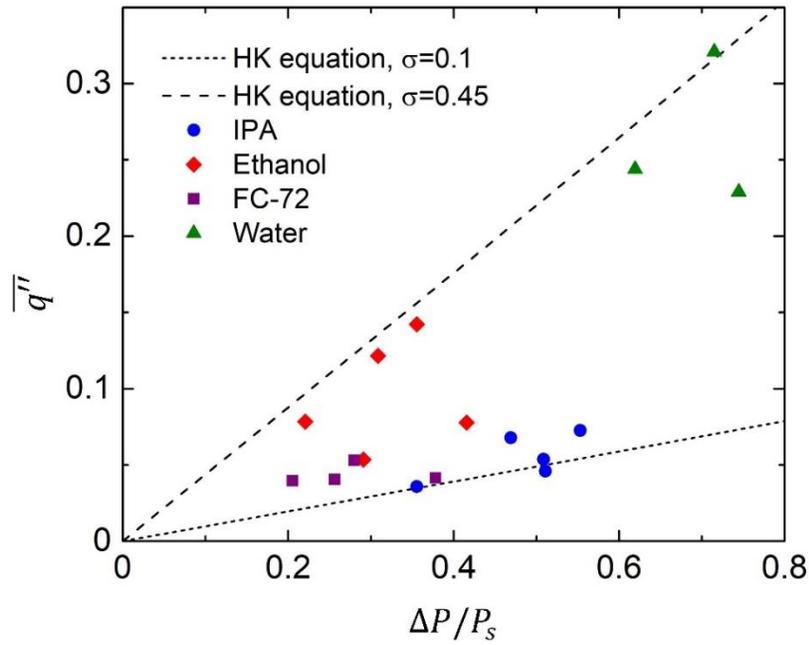

*Figure 4: Normalized evaporative heat flux ($\overline{q''}$, defined as $q''/(\rho_v h_{fg}\sqrt{\frac{RT_s}{2\pi}})$) for different fluids at the onset of pore-level evaporation regime shown in Figure 2 as a function of the dimensionless driving potential $\Delta P/P_s$ ($\Delta P = P_s - P_V$, where $P_s$ is the saturated pressure at $T_s$ and $P_V$ is the vapor pressure). The two dashed lines represents the kinetic limit results calculated from the HK equation with different values for the accommodation coefficient (σ = 0.1 and 0.45).*

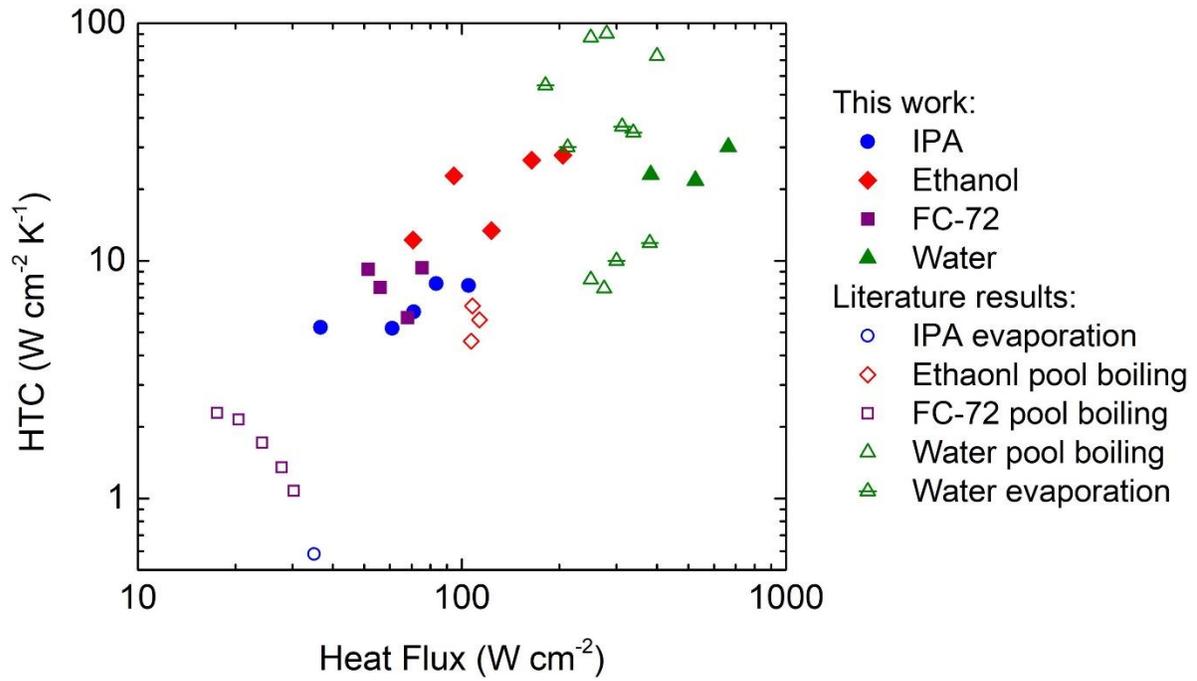

*Figure 5: HTC-heat flux plot at the transition points for different fluids. PCHT experiments for these fluids reported in literature are also plotted for comparison, including evaporation of IPA [49] and water [36, 38], and pool boiling of ethanol [46], FC-72 [48], and water [9, 13, 50].*

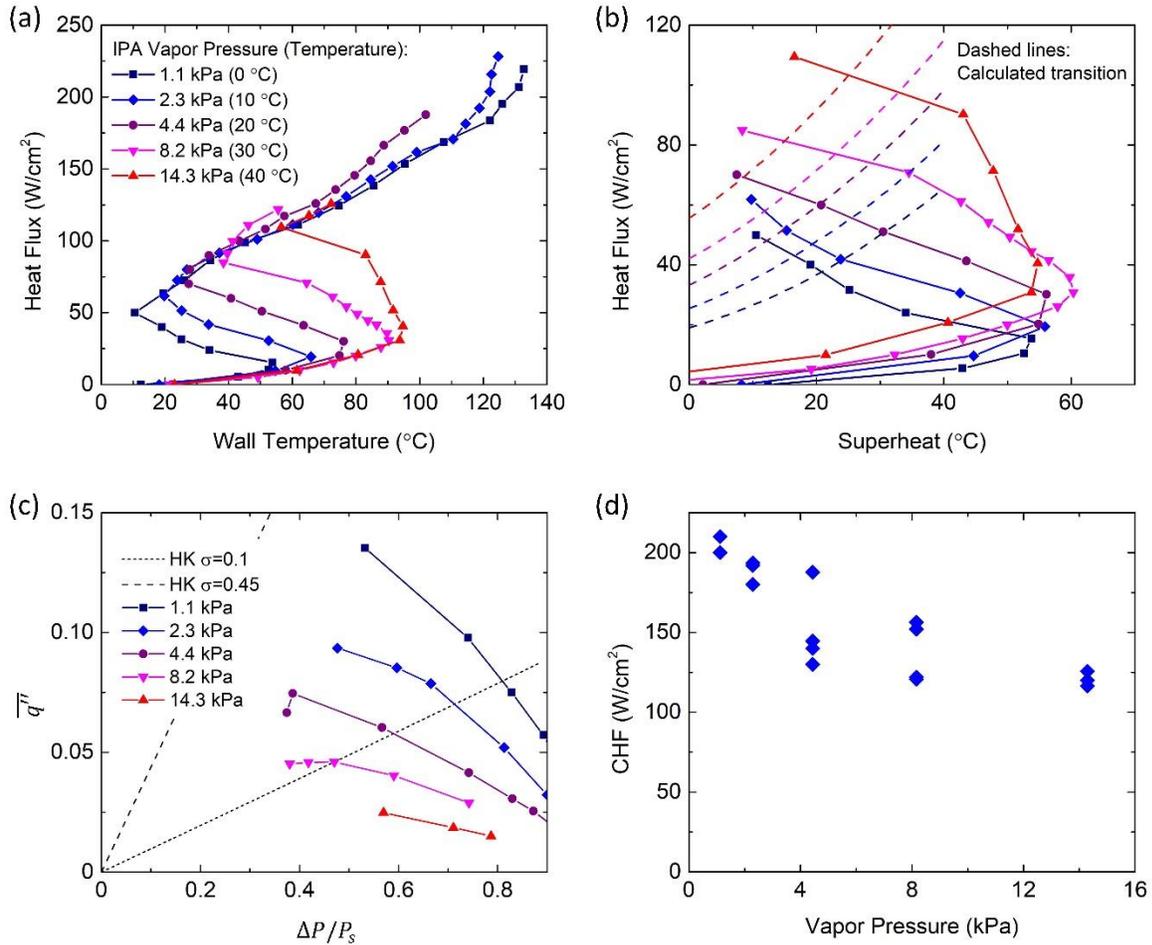

*Figure 6: (a) Heat flux as a function of wall temperature with varying vapor pressure and similar pressure difference (ranging from 93.1 to 100.2 kPa) using IPA as the working fluid. For vapor pressure of 1.1, 2.3, 4.4, and 8.2 kPa, the liquid pressure was 101.3 kPa; for vapor pressure of 14.3 kPa, the liquid pressure was 111.3 kPa. (b) Heat flux as a function of superheat for the thin film boiling part of the curves shown in (a), along with the model for transition (dashed lines). The color of the dashed lines corresponds to the experimental conditions of the curves with the same color shown in (a). (c) Dimensionless heat flux as a function of the dimensionless driving potential for the evaporation part of the curves shown in (a), along with the HK models. (d) Experimental CHF of IPA under different vapor pressure conditions.*

# Supplementary Information

## Transition between Thin Film Boiling and Evaporation on Nanoporous Membranes Near the Kinetic Limit

Qingyang Wang [a], Yang Shi [a,b], Renkun Chen [a,*]

[a] Department of Mechanical and Aerospace Engineering, University of California, San Diego, La Jolla, California 92093-0411, United States

[b] School of Mechanical Engineering and Automation, Harbin Institute of Technology (Shenzhen), Shenzhen 518000, China

*Email: rkchen@ucsd.edu

## S1: Fluid temperature inside the membrane

Since the sensible heat $c_p(T_s - T_L)$ of the liquid under the experimental conditions is at least an order of magnitude smaller than the latent heat $h_{fg}$, the Reynolds number of liquid flow inside the pore can be roughly estimated by

$$\text{Re}_{D_h} = \frac{q'' D_h}{\eta \mu h_{fg}}$$

where $\mu$ is the viscosity of the fluid, $\eta$ is the membrane porosity, $q''$ is the applied heat flux, and $D_h$ is the hydraulic diameter of the nanopores, as similarly defined in Eq. (1) in the manuscript. Under the experimental conditions, the calculated Reynolds number is on the order of $10^{-3}$ or smaller, which indicates that the convective heat transfer of liquid flow inside the pores can be negligible and the heat transfer across the membrane can be considered as pure conduction. The thermal resistance network is shown in Figure S1.

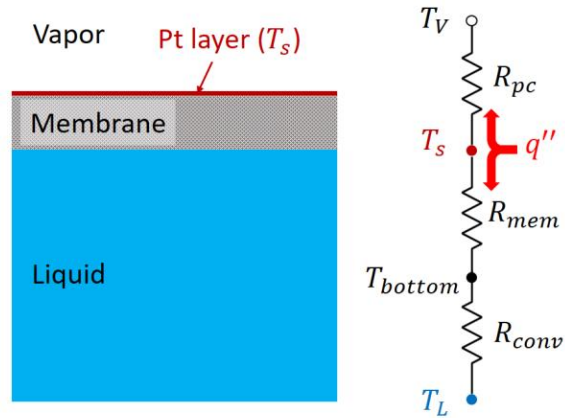

*Figure S1: Schematic and the thermal circuit of the membrane configuration.*

Here we demonstrate the calculation using the experiment of ethanol with 101.3 kPa liquid pressure (the purple curve shown in Fig. 2a in the main manuscript) as an example. The superheat close to the ending of thin film boiling is ~10 °C. The convective heat transfer coefficient between the membrane and the bottom liquid bath below the membrane surface can be estimated using the correlation for natural convection of a heated horizontal plate facing down as [1]

$$h = \frac{k}{L}\text{Nu}_L = \frac{k}{L}0.82\text{Ra}_L^{1/5}$$

where $k$ is the thermal conductivity of the liquid. With the thermophysical properties of ethanol, the heat transfer coefficient $h = 238 \text{ W m}^{-2}\text{K}^{-1}$, which gives a thermal resistance $R_{conv} =$

$4.2 \times 10^{-3}$ m²K W⁻¹. Note that the heat transfer coefficient is overestimated, since the correlation deals with free space convection while the actual liquid bath below the membrane is constrained by the liquid feeding column. Still, this convective thermal resistance is significantly larger than the overall thermal resistance of the system ($\sim 10^{-5}$ m²K W⁻¹), which indicates that the phase change thermal resistance $R_{pc}$ is close to the overall thermal resistance. Therefore, the heat flux flowing along the thermal pathway on the downward direction in Fig. S1 is at most ~0.24% of the total heat flux $q''$ supplied to the heater. The membrane thermal resistance $R_{mem}$ can be estimated by $t_{mem}/k_{mem}$ where $k_{mem}$ is the effective thermal conductivity of the liquid-filled membrane. Thus, the upper bound of the temperature drop across the membrane $T_s - T_{bottom}$ is 0.24% · $q''t_{mem}/k_{mem}$. We further conservatively assume that $k_{mem}$ is as low as that for an air-filled membrane (close to the membrane itself) which has cross-plane thermal conductivity of ~1.5 W m⁻¹ K⁻¹ [2], which leads to an upper bound of only ~0.1 °C. This small temperature difference indicates that the liquid flowing along the pores has temperature very close to the measured wall temperature, which justifies our choice of using the wall temperature as the reference temperature for fluid viscosity.

**S2: Estimated capillary pressure**

For the low surface tension fluid tested in this work, thin film evaporation regime is realized in the experiments, and thus, the CHF of these experiments are higher than the maximum heat flux in thin film boiling calculated using Eq. (1). We can estimate the capillary pressure at the highest heat flux using the following equation

$$q''_{CHF} = \frac{D_h^2 (P_L - P_V + P_c)}{32 \mu L} \rho_l [h_{fg} + c_p(T_s - T_L)] \eta$$

where $P_c$ is the capillary pressure providing extra liquid supply. For an exemplary curve of ethanol with 101.3 kPa liquid pressure shown in Fig. 2a in the main manuscript, CHF of 196.2 W cm⁻² was achieved at a superheat of 26.9 °C. Using this equation, the capillary pressure at CHF is ~45 kPa, which is much smaller than ~331 kPa estimated by the Young-Laplace equation as $P_c = 2\gamma/r_p$ assuming zero contact angle for ethanol on alumina, where $\gamma$ is the surface tension of ethanol and $r_p$ is the radius of the nanopore. Therefore, the evaporation CHFs recorded in the experiments are still below theoretical values.

## S3: Experimental setup

The schematic of the experimental setup used in this work is shown in Figure S2a. Two chambers were used each with a controlled pressure in a specific experiment. The pressure in the liquid chamber, $P_L$, was fixed at a preset value before and during each experiment by adjusting the air pressure inside the chamber with either an air compressor (for above-atmospheric pressure) or a vacuum pump (for sub-atmospheric pressure). The pressure in the vapor chamber ($P_V$) was constantly monitored using a digital pressure gauge and maintained at the preset values using an adjustable valve and a vacuum pump with a PID control loop. The small liquid flow rate in the tubing connecting the two chambers in the experiments produces negligible pressure drop. Therefore, the liquid pressure underneath the membrane is approximately the same as the liquid chamber pressure, $P_L$. Figure S2 b-c shows the schematic and the photo of the sample mounted on the Acrylic sample stage inside the vapor chamber. To make good electrical contacts to the Cu pads, thick Sn foils were pressed against the contact pads using custom-made PMMA clamps and set screws, as shown in Figure S2c. Figure S2d shows the procedure to assemble the sample. Figure S2e shows the photo of the vapor chamber consisting of an aluminum chamber (to collect overflowed liquid) and a glass cover (to visualize the experiments). The electrical feedthrough into the vapor chamber is on the back side of the aluminum chamber and connected to the computer.

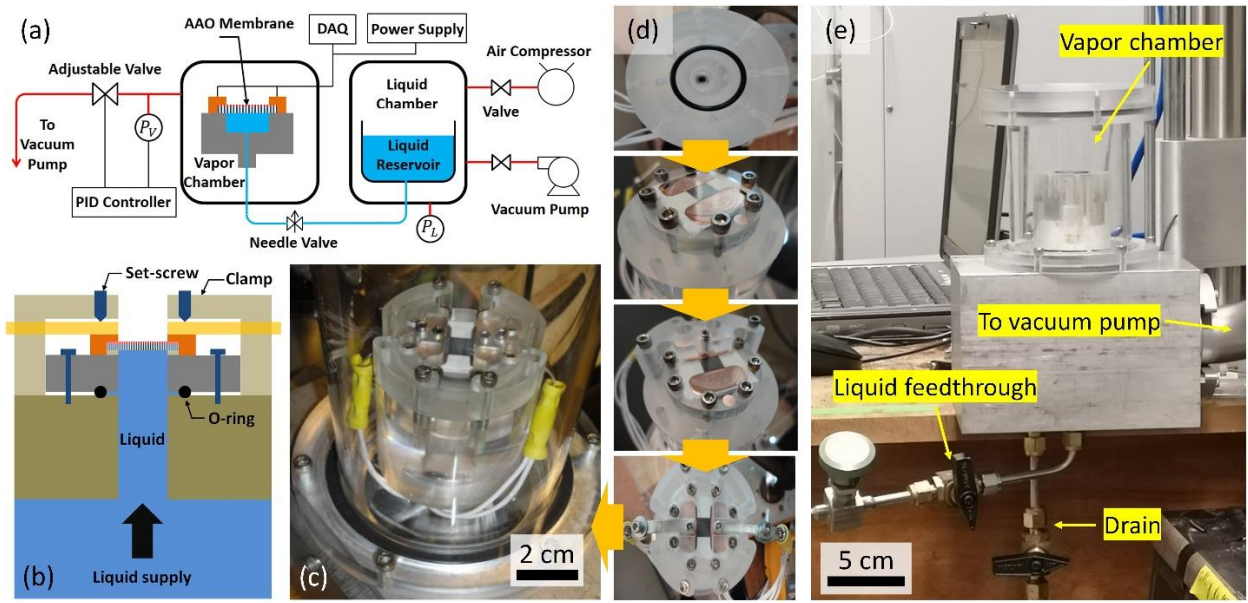

*Figure S2: (a) Schematic of the experimental setup. (b) Schematic of the sample mounting procedure. (c) Photo of one sample mounted for experiment. (d) Procedure of assembling one sample into the vapor chamber shown in (c). (e) Photo of the vapor chamber without sample.*

## S4: Kinetic limit calculation.

The kinetically limited interfacial heat flux is calculated using the Hertz-Knudsen equation as [3]

$$q_k'' = h_{fg}\dot{m} = \sigma h_{fg} \frac{1}{\sqrt{2\pi R}} \left( \frac{P_{eq}}{\sqrt{T_i}} - \frac{P_V}{\sqrt{T_V}} \right)$$

where the interface temperature $T_i$ is taken as the membrane surface temperature $T_s$, and the equilibrium pressure $P_{eq}$ is taken as the saturation pressure at $T_s$. Due to the small temperature difference, the dimensionless kinetic limited heat flux is almost proportional to the dimensionless driving potential $\Delta P/P_s$ and fluid-independent, as seen in Figures 4 and 6c in the main manuscript.

## S5: Images of the sample surface

Figure S3 shows the images of the heated sample surface for ethanol experiments with 101.3 kPa liquid pressure and 5.9 kPa vapor pressure. Figure S3a corresponds to the pool boiling regime where the bubbles are ~mm in size. Figure S3b corresponds to the thin film boiling regime where the bubble size is much smaller due to the constrain by the liquid film thickness. Figure S3c corresponds to the transition points where no bubbles can be observed, and evaporation became the main heat transfer mode. Similar images for this experiment were also published previously in our previous work [4].

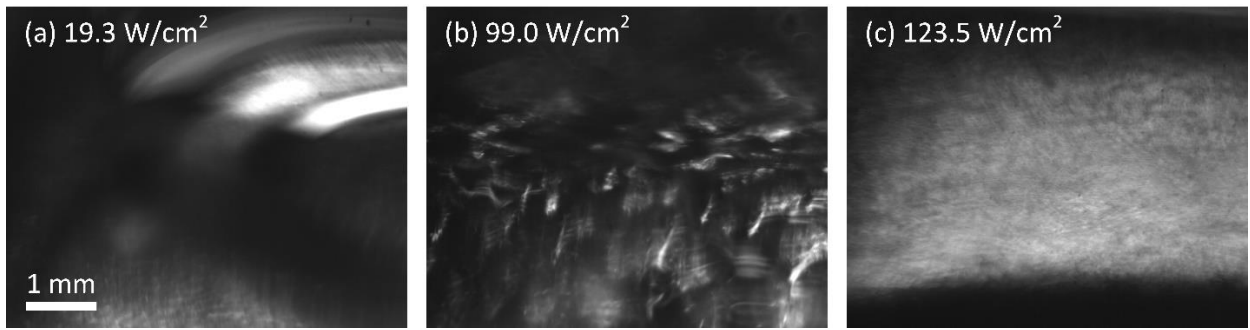

*Figure S3: High speed images of the heated surface during the experiments with 101.3 kPa liquid pressure and 5.9 kPa vapor pressure, using ethanol as the working fluid, at various heat fluxes: (a) 19.3 W cm⁻²; (b) 99.0 W cm⁻²; (c) 123.5 W cm⁻².*

## S6: Heat flux-superheat plot for kinetic limit

In Figure 6c of the main manuscript, we plotted the dimensionless heat flux as a function of the dimensionless driving potential for the evaporation part of the curves shown in Figure 6a, and compared with the HK models with accommodation coefficient of 0.1 and 0.45, respectively. In Figure S4 below, the HK models and the experimental curves are also plotted together in heat flux-superheat plot. Figure S4a is the exact same plot as Figure 6c, while Figure S4 b-f each shows the experimental curve for one vapor pressure condition along with two HK models under that condition. For different vapor pressure, the HK models with the same accommodation coefficient produce different heat flux under the same superheat, but they collapse into one curve in the dimensionless plot as shown in Figure S4a.

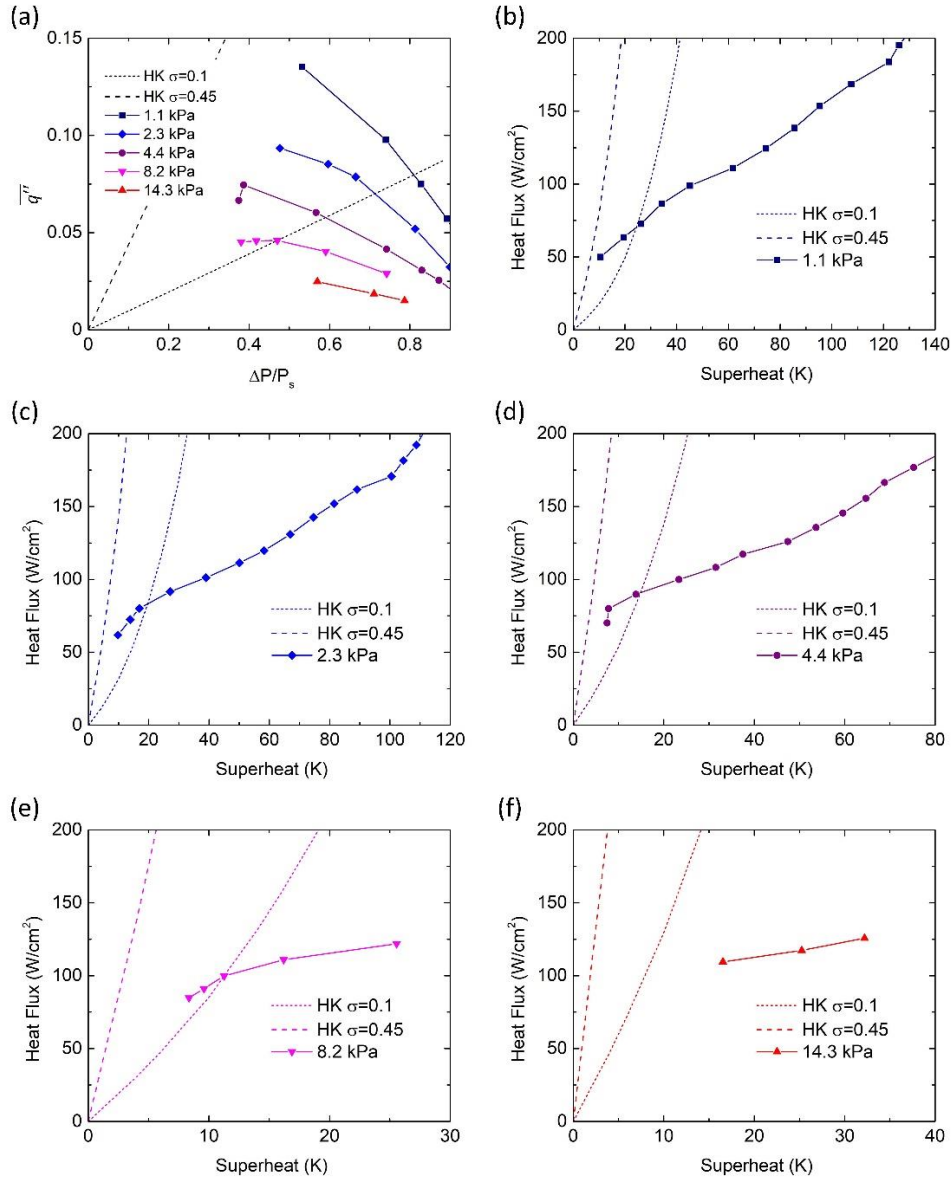

*Figure S4:(a) The same as Figure 6c, showing the dimensionless heat flux vs. dimensionless driving potential for the evaporation part of the curves shown in Figure 6a. (b-f) Heat flux vs. superheat curves along with HK model calculation results for the curves shown in (a), where each panel represents one vapor pressure condition.*

## Supplementary References

[1] A.F. Mills, C. Coimbra, Basic heat transfer, Temporal Publishing, LLC, 2015.
[2] T. Borca-Tasciuc, A. Kumar, G. Chen, Data reduction in 3ω method for thin-film thermal conductivity determination, Review of scientific instruments, 72(4) (2001) 2139-2147.
[3] V.P. Carey, Liquid-vapor Phase-change Phenomena: An Introduction to the Thermophysics of Vaporization and Condensation Processes in Heat Transfert Equipment, Hemisphere publishing corporation, 1992.
[4] Q. Wang, R. Chen, Widely tunable thin film boiling heat transfer through nanoporous membranes, Nano Energy, 54 (2018) 297-303.